\documentclass[RNAAS]{aastex62}

\usepackage{graphicx}
\usepackage{graphics}

\begin{document}

\title{3DMAP-VR, a project to visualize 3-dimensional models of
astrophysical phenomena in virtual reality}

\correspondingauthor{Salvatore Orlando}
\email{salvatore.orlando@inaf.it}

\author[0000-0003-2836-540X]{Salvatore Orlando}
\affiliation{INAF - Osservatorio Astronomico di Palermo ``G.S.
Vaiana'', Piazza del Parlamento 1, 90134 Palermo, Italy}

\author[0000-0003-4948-6550]{Ignazio Pillitteri}
\affiliation{INAF - Osservatorio Astronomico di Palermo ``G.S.  
Vaiana'', Piazza del Parlamento 1, 90134 Palermo, Italy}

\author[0000-0002-2321-5616]{Fabrizio Bocchino}
\affiliation{INAF - Osservatorio Astronomico di Palermo ``G.S.
Vaiana'', Piazza del Parlamento 1, 90134 Palermo, Italy}

\author[0000-0002-0640-081X]{Laura Daricello}
\affiliation{INAF - Osservatorio Astronomico di Palermo ``G.S.
Vaiana'', Piazza del Parlamento 1, 90134 Palermo, Italy}

\author[0000-0002-7667-0479]{Laura Leonardi}
\affiliation{INAF - Osservatorio Astronomico di Palermo ``G.S.
Vaiana'', Piazza del Parlamento 1, 90134 Palermo, Italy}


\section{} 

Fully 3D magnetohydrodynamic (MHD) simulations of astrophysical
phenomena represent a challenge in standard data visualization for
scientific purposes, for the amount of processed data and the wealth
of scientific information they contain. Recently, the potential of
virtual reality (VR) hardware and software began to be exploited for
the purposes of scientific data analysis. Moreover, VR has also
been adopted in different fields of public outreach and education
with excellent outcome. To this end, YouTube and online multimedia
digital stores host several VR titles with high visual impact in
the categories of Astrophysics and Space Science. However, the
routinely scientific use of the VR environment is still in its
infancy and requires the development of ad-hoc techniques and
methods.

In the first half of 2019, we launched
\href{http://cerere.astropa.unipa.it/progetti_ricerca/HPC/3dmap_vr.htm}{3DMAP-VR}
(3-Dimensional Modeling of Astrophysical Phenomena in Virtual
Reality), a project aimed at visualizing 3D MHD models of astrophysical
simulations, using VR sets of equipment. The models account for all the
relevant physical processes in astrophysical phenomena: gravity,
magnetic-field-oriented thermal conduction, energy losses due to
radiation, gas viscosity, deviations from proton-electron temperature
equilibration, deviations from the ionization equilibrium, cosmic
rays acceleration, etc. (e.g. \citealt{2011MNRAS.415.3380O,
2015ApJ...810..168O, 2016ApJ...822...22O, 2017MNRAS.464.5003O}).

The workflow to create VR visualizations of the models combines:
1) accurate 3D HD/MHD simulations performed for scientific purposes,
using parallel numerical codes for astrophysical plasmas (e.g. the
FLASH code, \citealt{for00}, or the PLUTO code, \citealt{2007ApJS..170..228M})
on high performance computing facilities (e.g. CINECA, Bologna,
Italy); 2) data analysis and visualization applications (e.g.
\href{https://www.harrisgeospatial.com/Software-Technology/IDL}{Interactive
Data Language}, \href{https://yt-project.org}{YT project},
\href{https://www.paraview.org}{ParaView},
\href{https://wci.llnl.gov/simulation/computer-codes/visit/}{Visit},
\href{http://www.meshlab.net}{MeshLab},
\href{http://www.meshmixer.com}{MeshMixer}) to realize navigable
3D graphics of the astrophysical simulations and quickly have a VR
representation of the models. The 3D representations are realized
using a mixed technique consisting of multilayer isodensity surfaces
with different opacities. Once the 3D graphics are ready, they are
uploaded on \href{https://sketchfab.com}{Sketchfab}, one of the
largest open access platforms to publish and share 3D virtual reality
and augmented reality content. Our VR laboratory includes two Oculus
Rift VR sets of equipment and dedicated computers with advanced
graphics cards to visualize the models in VR. The laboratory is
used to analyze the numerical results in an immersive fashion,
integrating the traditional screen displays, and allows scientists
to navigate and interact with their own MHD models. At the same
time, we use the VR in educational and public outreach events in
order to visualize invisible radiation, improve the learners sense
of presence and, eventually, increase the content understanding and
motivation to learn.

\begin{figure}[th!]
\begin{center}
\includegraphics[width=18cm]{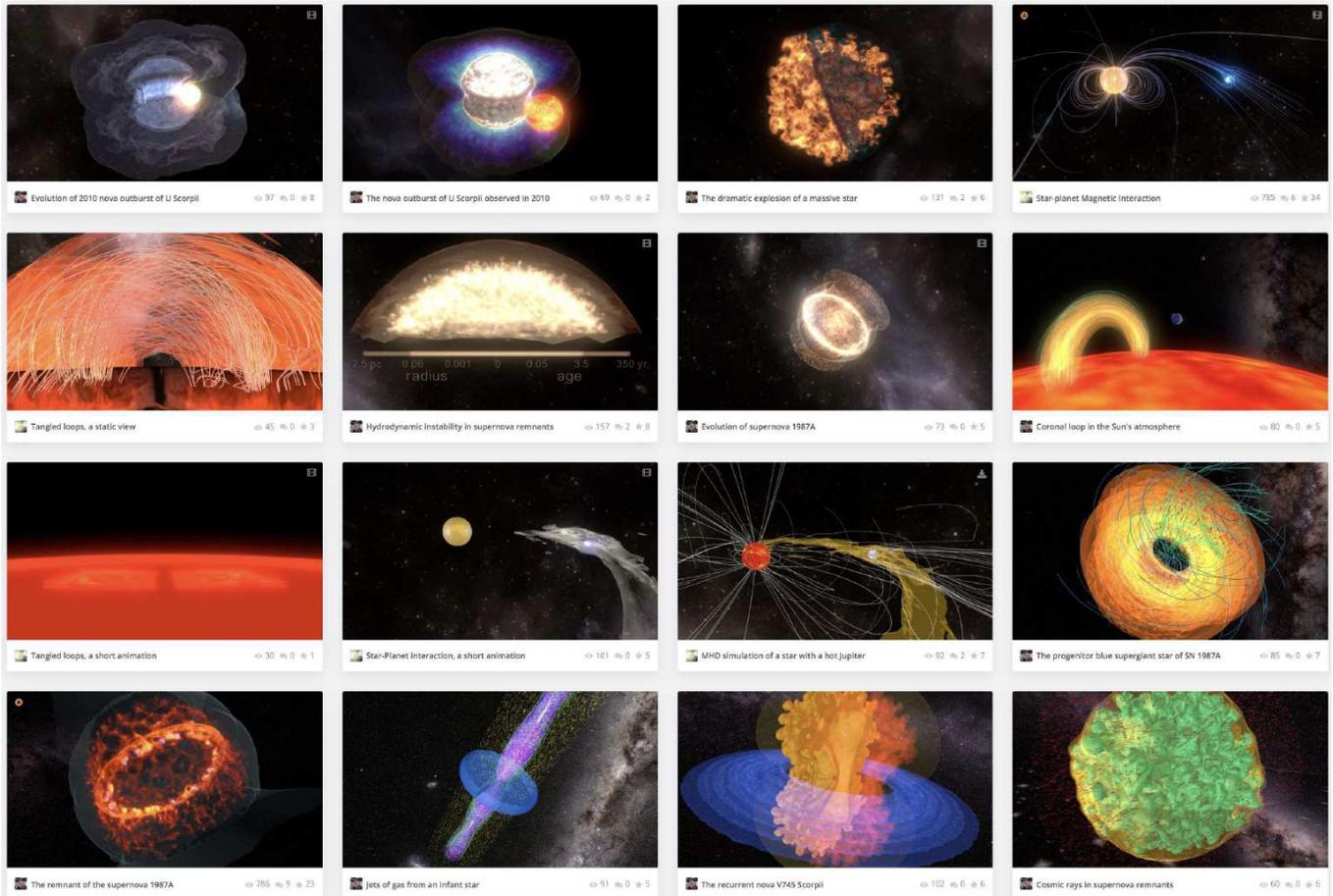}
\caption{Examples of 3D models uploaded in the Sketchfab gallery
\href{https://skfb.ly/6NooE}{"Universe in hands"}.\label{fig:1}}
\end{center}
\end{figure}

We realized an excellent synergy between our 3DMAP-VR project and
Sketchfab to promote a wide dissemination of results for both
scientific and public outreach purposes. We realized a Sketchfab
gallery, \href{https://skfb.ly/6NooE}{"Universe in hands"}, which
gathers different models of astrophysical objects and phenomena
developed by our team for scientific purposes and published in
international scientific journals. More specifically these models
describe (see Fig.~\ref{fig:1}): magnetic structures of the
solar and stellar coronae (e.g. \citealt{2016ApJ...830...21R}); the
interaction between a star and its planet (cf.
\citealt{2015ApJ...805...52P}); accretion phenomena in young stellar
objects (e.g. \citealt{2011MNRAS.415.3380O}); protostellar jets
(e.g \citealt{2016A&A...596A..99U}); nova outbursts (e.g.
\citealt{2010ApJ...720L.195D}); the outcome of supernova explosions
(e.g. \citealt{2016ApJ...822...22O}); the interaction of supernova
remnants with the inhomogeneous surrounding environment (e.g.
\citealt{2015ApJ...810..168O}); the effects of cosmic ray particle
acceleration on the morphology of supernova remnants (e.g.
\citealt{2012ApJ...749..156O}). In addition we created a second
gallery, \href{https://skfb.ly/6OzCU}{"The art of astrophysical
phenomena"}, which collects artist's view of astrophysical phenomena
for public outreach purposes. The two galleries are publically
available and continuously updated to include new models.

\begin{thebibliography}{}
\expandafter\ifx\csname natexlab\endcsname\relax\def\natexlab#1{#1}\fi
\providecommand{\url}[1]{\href{#1}{#1}}
\providecommand{\dodoi}[1]{doi:~\href{http://doi.org/#1}{\nolinkurl{#1}}}
\providecommand{\doeprint}[1]{\href{http://ascl.net/#1}{\nolinkurl{http://ascl.net/#1}}}
\providecommand{\doarXiv}[1]{\href{https://arxiv.org/abs/#1}{\nolinkurl{https://arxiv.org/abs/#1}}}

\bibitem[{{Drake} \& {Orlando}(2010)}]{2010ApJ...720L.195D}
{Drake}, J.~J., \& {Orlando}, S. 2010, \apjl, 720, L195,
  \dodoi{10.1088/2041-8205/720/2/L195}

\bibitem[{{Fryxell} {et~al.}(2000){Fryxell}, {Olson}, {Ricker}, {Timmes},
  {Zingale}, {Lamb}, {MacNeice}, {Rosner}, {Truran}, \& {Tufo}}]{for00}
{Fryxell}, B., {Olson}, K., {Ricker}, P., {et~al.} 2000, \apjs, 131, 273

\bibitem[{{Mignone} {et~al.}(2007){Mignone}, {Bodo}, {Massaglia}, {Matsakos},
  {Tesileanu}, {Zanni}, \& {Ferrari}}]{2007ApJS..170..228M}
{Mignone}, A., {Bodo}, G., {Massaglia}, S., {et~al.} 2007, \apjs, 170, 228,
  \dodoi{10.1086/513316}

\bibitem[{{Orlando} {et~al.}(2012){Orlando}, {Bocchino}, {Miceli}, {Petruk}, \&
  {Pumo}}]{2012ApJ...749..156O}
{Orlando}, S., {Bocchino}, F., {Miceli}, M., {Petruk}, O., \& {Pumo}, M.~L.
  2012, \apj, 749, 156, \dodoi{10.1088/0004-637X/749/2/156}

\bibitem[{{Orlando} {et~al.}(2017){Orlando}, {Drake}, \&
  {Miceli}}]{2017MNRAS.464.5003O}
{Orlando}, S., {Drake}, J.~J., \& {Miceli}, M. 2017, \mnras, 464, 5003,
  \dodoi{10.1093/mnras/stw2718}

\bibitem[{{Orlando} {et~al.}(2015){Orlando}, {Miceli}, {Pumo}, \&
  {Bocchino}}]{2015ApJ...810..168O}
{Orlando}, S., {Miceli}, M., {Pumo}, M.~L., \& {Bocchino}, F. 2015, \apj, 810,
  168, \dodoi{10.1088/0004-637X/810/2/168}

\bibitem[{{Orlando} {et~al.}(2016){Orlando}, {Miceli}, {Pumo}, \&
  {Bocchino}}]{2016ApJ...822...22O}
---. 2016, \apj, 822, 22, \dodoi{10.3847/0004-637X/822/1/22}

\bibitem[{{Orlando} {et~al.}(2011){Orlando}, {Reale}, {Peres}, \&
  {Mignone}}]{2011MNRAS.415.3380O}
{Orlando}, S., {Reale}, F., {Peres}, G., \& {Mignone}, A. 2011, \mnras, 415,
  3380, \dodoi{10.1111/j.1365-2966.2011.18954.x}

\bibitem[{{Pillitteri} {et~al.}(2015){Pillitteri}, {Maggio}, {Micela},
  {Sciortino}, {Wolk}, \& {Matsakos}}]{2015ApJ...805...52P}
{Pillitteri}, I., {Maggio}, A., {Micela}, G., {et~al.} 2015, \apj, 805, 52,
  \dodoi{10.1088/0004-637X/805/1/52}

\bibitem[{{Reale} {et~al.}(2016){Reale}, {Orlando}, {Guarrasi}, {Mignone},
  {Peres}, {Hood}, \& {Priest}}]{2016ApJ...830...21R}
{Reale}, F., {Orlando}, S., {Guarrasi}, M., {et~al.} 2016, \apj, 830, 21,
  \dodoi{10.3847/0004-637X/830/1/21}

\bibitem[{{Ustamujic} {et~al.}(2016){Ustamujic}, {Orlando}, {Bonito}, {Miceli},
  {G{\'o}mez de Castro}, \& {L{\'o}pez-Santiago}}]{2016A&A...596A..99U}
{Ustamujic}, S., {Orlando}, S., {Bonito}, R., {et~al.} 2016, \aap, 596, A99,
  \dodoi{10.1051/0004-6361/201628712}

\end{thebibliography}

\end{document}